\documentclass[
preprint,
 amsmath,amssymb,
 aps,
superscriptaddress
]{revtex4-2}
\usepackage{amsmath}
\usepackage{graphicx}
\usepackage{dcolumn}
\usepackage{lipsum}
\usepackage{bm}

\usepackage[dvipsnames]{xcolor}
\usepackage{appendix}
\begin{document}

\preprint{APS/123-QED}

\title{Thermally-generated spin current in the topological insulator Bi$_2$Se$_3$
}
\author{Rakshit Jain}
\affiliation{Department of Physics, Cornell University. Ithaca, NY 14853, USA}
\affiliation{ School of Applied and Engineering Physics, Cornell University. Ithaca, NY 14853, USA}
\author{Max Stanley}
\affiliation{Department of Physics and Materials Research Institute, The Pennsylvania State University, University Park, 16802, Pennsylvania, USA}
\author{Arnab Bose}
\affiliation{Department of Physics, Cornell University. Ithaca, NY 14853, USA}
\affiliation{ School of Applied and Engineering Physics, Cornell University. Ithaca, NY 14853, USA}
\author{Anthony R. Richardella}
\affiliation{Department of Physics and Materials Research Institute, The Pennsylvania State University, University Park, 16802, Pennsylvania, USA}
\author{Xiyue S. Zhang}
\affiliation{ School of Applied and Engineering Physics, Cornell University. Ithaca, NY 14853, USA}
\author{Timothy Pillsbury}
\affiliation{Department of Physics and Materials Research Institute, The Pennsylvania State University, University Park, 16802, Pennsylvania, USA}
\author{David A. Muller}
\affiliation{ School of Applied and Engineering Physics, Cornell University. Ithaca, NY 14853, USA}
\affiliation{Kavli Institute at Cornell for NanoScale Science, Ithaca, NY 14853, USA}
\author{Nitin Samarth}
\affiliation{Department of Physics and Materials Research Institute, The Pennsylvania State University, University Park, 16802, Pennsylvania, USA}
\author{Daniel C. Ralph}
\affiliation{Department of Physics, Cornell University. Ithaca, NY 14853, USA}
\affiliation{Kavli Institute at Cornell for NanoScale Science, Ithaca, NY 14853, USA}
\email{dcr14@cornell.edu}

\date{\today}
\begin{abstract}

We complete measurements of interconversions among the full triad of thermal gradients, charge currents, and spin currents in the topological insulator Bi$_2$Se$_3$ by quantifying the efficiency with which thermal gradients can generate transverse spin currents.  We accomplish this by
comparing the spin Nernst magneto-thermopower to the spin Hall magnesistance for bilayers of Bi$_2$Se$_3$/CoFeB.  We find that Bi$_2$Se$_3$ does generate substantial thermally-driven spin currents.  A lower bound for the ratio of spin current to thermal gradient is $J_s/\nabla_x T$ =  (4.9 $\pm$ 0.9)  $\times$ 10$^{6}$ ($\hbar/2e$)  A~m$^{-2}$ / K $\mu$m$^{-1}$, and a lower bound for the magnitude of the spin Nernst ratio is $-$0.61 $\pm$ 0.11. The spin Nernst ratio for Bi$_2$Se$_3$ is the largest among all  materials measured to date, 2-3 times larger compared to previous measurements for the heavy metals Pt and W.

\end{abstract}

\maketitle
\section{Introduction}
Topological insulators (TIs) provide the most-efficient known transduction between charge current density and spin current density (i.e., the spin Hall effect)~\cite{mellnik2014spin, fan2014magnetization, wu2019room, jamali2015giant, wang2016surface}, thereby producing spin-orbit torques capable of driving magnetization switching in magnetic memory structures ~\cite{han2017room, dc2018room, khang2018conductive}.  In addition, TIs can also very efficiently transduce thermal gradients to electric field via the Seebeck effect~\cite{vineis2010nanostructured, hor2009p, guo2016tuning}, with potential for thermoelectric applications.
Here, for the first time we measure the efficiency of transduction for all three legs of the triad between thermal gradients, charge currents, and spin currents for a topological insulator/magnet bilayer (see Fig.~1(b)).
In particular, we provide the first measurement of the efficiency by which a thin film of the topological insulator Bi$_2$Se$_3$ can transduce a thermal gradient to spin current.  
Understanding thermally-generated spin currents in topological insulators is important for characterizing the effect of Joule heating on measurements of current-induced spin-orbit torques~\cite{roschewsky2019spin}.  If the thermal spin currents are sufficiently strong, they could in principle be put to use in generating useful torques~\cite{kim2020observation, bose2018control}.
We find that the magnitude of the spin Nernst ratio of Bi$_2$Se$_3$ is larger by a factor of 2-3 compared to previous reports for the heavy metals Pt~\cite{meyer2017observation, kim2017observation, bose2018direct, bose2019recent} and W~\cite{sheng2017spin, kim2017observation, bose2019recent}.

\section{Background} 

We measure thermally-generated spin currents using the same physics by which electrically-generated spin currents give rise to the spin Hall magnetoresistance (SMR) effect.  In the electrically-generated case, an electric field $E$ applied in the plane of a spin-source/ferromagnet bilayer gives rise to a vertically-flowing spin current density $J_s$ via the spin Hall effect, with an efficiency characterized by the spin Hall ratio, $\theta_{SH}$~\cite{liu2012spin}:
$J_s = \frac{\hbar}{2e}\frac{\theta_{SH}}{\rho_{SS}}E,$
where $\hbar$ is the reduced Planck constant, $e$ is the magnitude of the electron charge, and $\rho_{SS}$ is the electrical resistivity of the spin-source material. The degree of reflection of this spin current at the magnetic interface depends on the orientation of the magnetization $\hat{m}$ in the magnetic layer.  The reflected spin current produces a voltage signal by the inverse spin Hall effect, causing the resistance of the bilayer to depend on the magnetization angle~\cite{nakayama2013spin,chen2013theory, kim2016spin}:
$\Delta R(\hat{m}) = (1-m^2_y)\Delta R_{SMR}.$
\label{eq_SMR1}
Here $m_y$ is the component of the magnetization unit vector that is in-plane and perpendicular to the electric field.  
This resistance change corresponds to a voltage signal amplitude:
\begin{align}
\Delta V = \Delta R_{SMR} I = \frac{\Delta R_{SMR}}{R} E l =   \frac{2e}{\hbar} \frac{\Delta R_{SMR}}{R} \frac{\rho_{SS}}{\theta_{SH}} l J_s,
\label{eq_SMRvoltage}
\end{align}
with $I$  the total current through the bilayer, $R$  the total resistance of the bilayer, and $l$ the sample length.
Our analysis will assume that a thermally-generated spin current produces the same voltage signal as an electrically-generated spin current (i.e., that Eq.\ \ref{eq_SMRvoltage} holds for both cases with the same experimentally-measured value of $\Delta R_{SMR}/R$).  

An in-plane thermal gradient $\nabla_x T$ can similarly give rise to a vertically-flowing spin current in a spin-source layer via the spin Nernst effect~\cite{meyer2017observation, bose2019recent}.  Upon reflection of this spin current from a magnetic interface 
and then the action of the inverse spin Hall effect, this results in a 
voltage signal parallel to the thermal gradient that depends on the magnetization angle, in direct analogy to the spin Hall magnetoresistance. 
For experiments measured with an open-circuit electrical geometry (i.e., no net longitudinal charge flow in the bilayer) we will define the efficiency of spin current generation by the spin Nernst effect in terms of a spin Nernst parameter $\theta_{SN}$ 
\begin{align}
    J_s = -  \frac{\hbar}{2e}
    \frac{\theta_{SN}}{\rho_{SS}} S_{SS}\nabla_x T.
    \label{spincurrent_t}
\end{align}
Here $S_{SS}$ is the Seebeck coefficient of the spin-source layer and $\nabla_x T$ is the in-plane thermal gradient. The thermally-generated voltage takes the form~\cite{meyer2017observation, sheng2017spin, kim2017observation}
\begin{align}
    V^x_{th} = -l \nabla_x T (S + S_1 +  (1 - m_y^2)S_{SNT} ) 
    \label{vth_my}
\end{align}
where $S_{SNT}$ is the coefficient of the spin Nernst magneto-thermopower, $S_1$ is a magnetization-independent offset arising from the spin Nernst effect and inverse spin Hall effect, and  and $S$ denotes the total effective Seebeck coefficient of the bilayer given by 
$S = \frac{\rho_{SS} S_{FM}t_{FM} + \rho_{FM} S_{SS}t_{SS}}{\rho_{SS}t_{FM} + \rho_{FM}t_{SS} } \approx \chi S_{SS}.$
This approximation holds when the spin-source layer is a topological insulator such as Bi$_2$Se$_3$ for which the Seebeck coeeficient is much larger than for the ferromagnetic layer ($S_{SS} \gg S_{FM}$), and we define a current shunting ratio $\chi = {\rho_{FM}t_{SS}}/({\rho_{FM}t_{SS} + \rho_{SS}t_{FM}})$ where $\rho_{FM}$ is the resistivity of the ferromagnet and $t_{FM}$ and $t_{SS}$ are the thicknesses of the two layers.  As long as thermally-generated and electrically-generated spin currents are transduced to voltage the same way, we can combine Eqs.\ \ref{eq_SMRvoltage}-\ref{vth_my} to obtain
\begin{align}
\frac{J_s}{\nabla_x T} =  - S_{SNT} \frac{R}{\Delta R_{SMR}} \frac{\hbar}{2e} \frac{\theta_{SH}}{\rho_{SS}}
\label{eq_spin_current}
\end{align}
\begin{align}
\theta_{SN} =  \theta_{SH}  \frac{S_{SNT}}{S_{SS}} \frac{R}{\Delta R_{SMR}}.
\label{eq_S_SNT}
\end{align}
These are the two equations we will use to evaluate the thermally-generated spin current and the spin-Nernst ratio $\theta_{SN}$. 

For an open-circuit measurement, $\theta_{SN}$ will have contributions from both spin-current generated directly by a thermal gradient and spin current generated by an electric field that is also present due to the Seebeck effect.  It is therefore also of interest to separate these effects and define a spin current that would be generated by a thermal gradient alone, in the absence of any electric field, i.e., to define a ``bare'' spin Nernst ratio $\theta^0_{SN}$ such that
\begin{align}
J_s \equiv - \frac{\hbar}{2e} \theta_{SN}\frac{S_{SS}}{\rho_{SS}}  \nabla_x T &= \frac{\hbar}{2e} \frac{\theta_{SH}}{\rho_{SS}}E - \frac{\hbar}{2e} \frac{\theta^0_{SN}}{\rho_{SS}} S_{SS} \nabla_x T \\
&= \frac{\hbar}{2e} (\chi \theta_{SH} - \theta^0_{SN})\frac{S_{SS}}{\rho_{SS}} \nabla_x T
\label{eq_bareSN1}
\end{align}
Therefore, we can calculate
\begin{align}
    \theta^0_{SN} = \theta_{SN} + \chi \theta_{SH}. 
    \label{eq_bareSN2}
\end{align}

In our experimental geometry, a small vertical thermal gradient $\nabla_z T$ can also be present when we apply an in-plane thermal gradient.  This will produce additional background signals due to the ordinary Nernst effect (ONE), the spin Seebeck effect (SSE) + inverse spin Hall effect, and the anomalous Nernst effect (ANE):  
\begin{align}
    V_{th}^{z} = V_{ONE} \nabla_z T B_y + V_{SSE}  \nabla_z T m_y + V_{ANE} \nabla_z T m_z.
    \label{eq_OOP}
\end{align}
These signals will be distinguished from the voltages arising from an in-plane thermal gradient based on the different dependences on the magnetization orientation $\hat{m}$.

\section{Results}
We analyze bilayers of Bi$_2$Se$_3$ (8 nm)/ Co$_{20}$Fe$_{60}$B$_{20}$ ($\equiv$ CoFeB) (5 nm).  The 8 nm thickness of the Bi$_2$Se$_3$ was chosen to ensure negligible hybridization between states on the two surfaces~\cite{zhang2010crossover, neupane2014observation}. The Bi$_{2}$Se$_{3}$ thin films were grown by molecular beam epitaxy and initially capped with 20 nm of Se to protect them from air exposure while transporting them to a separate system for the deposition of the CoFeB. Details about the MBE growth can be found in the supplementary information.
High-quality growth of Bi$_2$Se$_3$ is confirmed by atomic force microscopy as well as x-ray diffraction measurements (see Fig.~S1).  The existence of a surface state on the Bi$_2$Se$_3$ thin films (with no Se cap or CoFeB overlayer) was also confirmed using angle resolved photoemission spectroscopy (ARPES), which measured a Dirac-like dispersion as shown in Fig.~1(c).  As is evident from the position of the Fermi-level in Fig.~1(c), the Bi$_2$Se$_3$ layer is electron-doped prior to the Se capping. This is consistent with previous studies which have identified the cause of the doping to be Se vacancies~\cite{hor2009p}.  After transfer to the separate vacuum system we heated the Bi$_2$Se$_3$/Se samples to a heater thermocouple temperature of 285 $^\circ$C for 3.5 hours to remove the Se cap.  We then deposited the CoFeB by DC magnetron sputtering followed by a 1.2 nm 
protective layer of Ta which forms TaO$_{x}$ upon air exposure. 
 Cross-sectional scanning transmission electron microscopy (Fig.~1(a)) shows that the bilayers possess a sharp interface with no visible oxidation at the interface.

\begin{figure}[h]
    \centering

    \includegraphics[width = \linewidth]{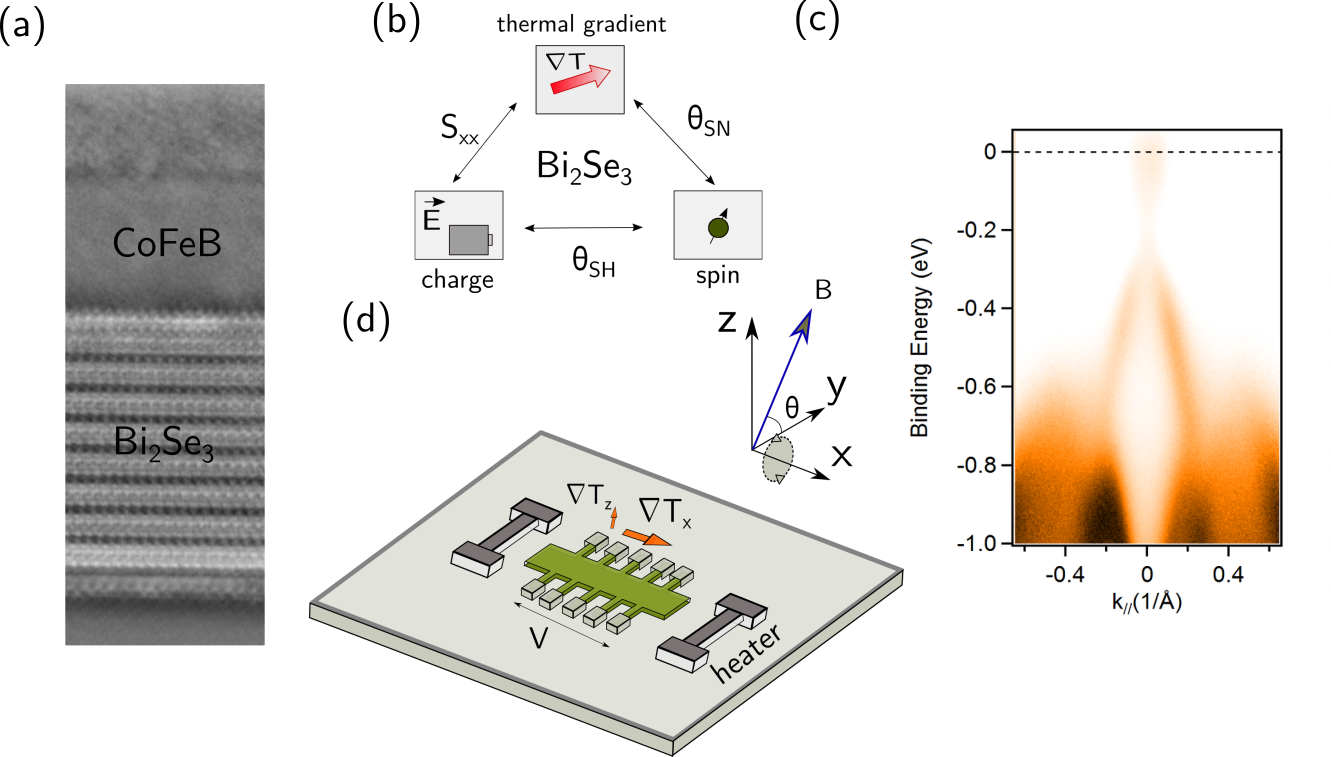}
    \caption{(a) Cross-Sectional scanning transmission electron microscope (STEM) image showing a sharp interface between Bi$_2$Se$_3$ and CoFeB. Bi$_2$Se$_3$ is crystalline, whereas CoFeB is amorphous. (b) Bi$_2$Se$_3$ 
    excels at the three figures of merit associated with interconversion between spin, charge and the thermal gradient  
    (c) ARPES measurements of the band structure of the top surface state of a single layer of Bi$_2$Se$_3$ of thickness 8 nm. The Fermi level is shown as the dotted line. k$_{\parallel}$ denotes the $\Gamma - M$ direction and the spectrum is obtained at room temperature. (d) Sample geometry for the spin Nernst magneto-thermopower experiments. 
    }
    \label{Fig1}
\end{figure}
\begin{figure}[h]
    \centering
    \includegraphics[width =  \linewidth]{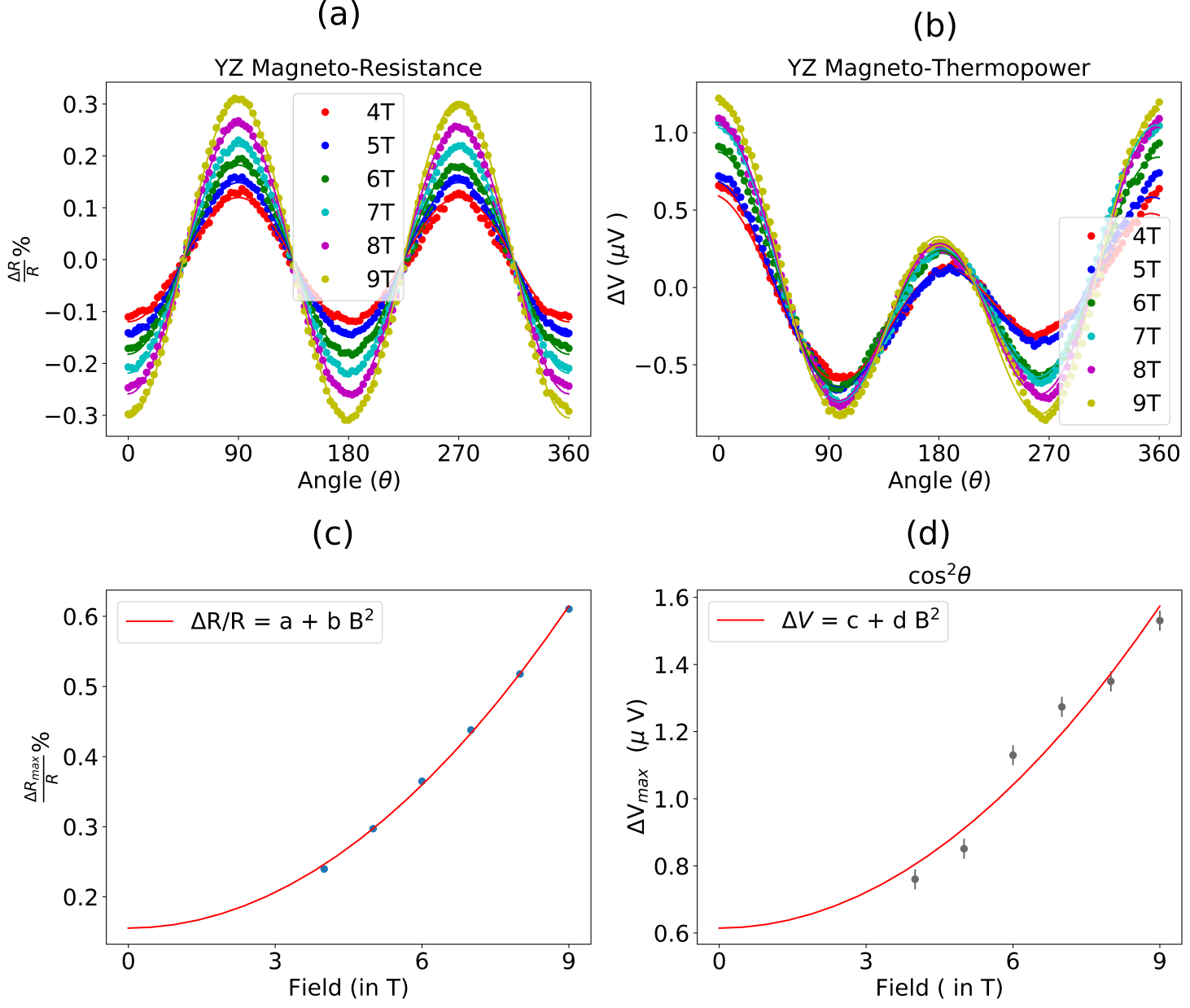}
    \caption{(a) Magnetoresistance percentage ratio ($\frac{\nabla R}{R} \times 100$) as a function of the magnetic field angle and magnitude for bilayers of Bi$_2$Se$_3$ (8 nm)/CoFeB (5 nm) at room temperature, for magnetic field rotated in the YZ plane. The four-point device resistance $R$ =  2.137 k$\Omega$.
    (b) YZ magneto-thermopower voltage as a function of the magnetic field angle and magnitude for the same sample. The heater power used for these sweeps was fixed at 442 mW (equivalent to a temperature drop of 6.8 K along the $l$ = 1.8 mm length of the device.)  
    (c) Amplitude of the YZ magnetoresistance as a function of magnetic-field magnitude, with a fit to the form $100 \times(a + b B^2)$.   The fit yields $a =0.155 \pm 0.05$ and $b = 0.0057\pm 0.0001$ T$^{-2}$.  
    (d) Field dependence of the cos$^2\theta$ part of YZ magneto-thermopower with a fit to the form $c + d B^2$. The fit yields $c =  0.61 \pm 0.07$  $\mu$V and $d = 0.012 \pm 0.001$  $\mu$V/T$^2$. 
    }
    \label{Fig2}
\end{figure}

We measure the spin current that is generated both electrically and thermally.  As a first step, we measure the spin Hall magnetoresistance of the bilayer.  We use optical lithography to pattern a Hall-bar sample geometry with 9 pairs of Hall contacts (only 5 are depicted in Fig.~1(d)) and make a 4-point measurement of the longitudinal resistance while rotating a magnetic field with fixed magnitude in the YZ  plane as defined by the diagram in Fig.~\ref{Fig1}(d).  Since the magnetization is always perpendicular to the current flow for this orientation of field sweep, this provides a measurement of the spin Hall magnetoresistance without contamination by the anisotropic magnetoresistance of the magnetic layer. Figure \ref{Fig2}(a) shows the magnetoresistance data for magnetic-field magnitudes of 4 to 9 Tesla, large enough  compared to the magnetic anisotropy (1.6 Tesla) that to a good approximation the magnetization is saturated along the field direction. 
The data fit well to the angular dependence $\Delta R/R = \Delta R_{max}(1-m^2_y) =\Delta R_{max} \cos^2\theta$.  The amplitude of the magnetoresistance for different magnitudes of magnetic field is plotted in  Fig.~\ref{Fig2}(c).  We find that the amplitude is well-described by the dependence $100 \times \Delta R_{max}/R = a + b B^2$, where $a$ and $b$ are constants.   

In measurements of separate samples, we find that the YZ magnetoresistance of both an individual Bi$_2$Se$_3$ layer (supplemental Fig.~S6(a)) and an individual CoFeB layer (Fig.~S6(c)) also have  $\cos^2\theta$ dependences. For the individual Bi$_2$Se$_3$ layer the amplitude of this signal is purely quadratic in $B$ with negligible zero-field offset (Fig.~S6(b)).  For the full bilayer, we therefore identify the contribution that is quadratic in magnetic field (Fig.~2(c)) with the magnetoresistance of the topological insulator and the field-independent component as due primarily to the spin Hall magnetoresistance. The YZ magnetoresistance of an individual CoFeB layer is weak, $100 \times \Delta R/R = 0.034 \pm 0.003$ (Fig.~S6(d)) but this is still about 1/5 the value for the full bilayer, so that we take it into account as a small correction to the primary signal (see supplemental material), 
Assuming that the magnetoresistance of the CoFeB layer and the SMR contribute in parallel to the sample conductance, 
 we estimate for the Bi$_2$Se$_3$/CoFeB bilayer that $100 \times \Delta R_{SMR}/R = 0.126~\pm~0.008$.  

We determine spin Nernst magneto thermopower $S_{SNT}$ by measuring the thermal analogue of YZ magnetoresistance, which henceforth we will refer to as the YZ magneto-thermopower.  We use the an experimental procedure similar to~\cite{bose2022origin}. We create a thermal gradient along the bilayer Hall bar using a lithographically-defined Pt heater adjacent to one end of the Hall bar, and measure the thermally-induced longitudinal voltage between the forth and last Hall contacts of the bilayer sample. 
Our samples contain two heaters, one near each end of the Hall bar, so that we can apply in-plane thermal gradients of either sign.  For each given heater power, we determine the temperature drop along the sample by measuring the local temperature at the position of the probed pairs of Hall contacts.  This is done by measuring the change in the 2-point resistance for each pair of Hall contacts and comparing to measurements of resistance versus temperature when an external heater is used to heat the full sample chip uniformly. 

The measured YZ magneto-thermopower of the Bi$_2$Se$_3$ bilayer for magnetic-field magnitudes from 4 to 9 Tesla is shown in Fig.~\ref{Fig2}(b). Figure 3(a,b) shows a representative analysis of the dependence on magnetic-field angle for two 4 Tesla 
 scans with opposite orientations of in-plane thermal gradients.
We observe two distinct contributions to the dependence of the thermopower on magnetic-field angle, $\propto \cos \theta$ and $\propto \cos^2 \theta$.  The $\cos \theta$ contribution can be understood as due to the terms arising for an out-of-plane component of the thermal gradient (Eq.~\ref{eq_OOP}). As expected, this contribution retains the same sign when the direction of the in-plane thermal gradient is reversed (compare Fig.~3(a) to Fig.~3(b)). 

The $\cos^2 \theta$ component within the YZ magneto-thermopower is our primary focus.  Like the magnetoresistance, this component of the magneto-thermopower can be fit to the form $\Delta S = c +d B^2$ with constants $c$ and $d$ -- i.e., it contains a contribution that depends on magnetic-field magnitude due to the Bi$_2$Se$_3$ by itself and a contribution that is independent of the magnetic field magnitude (Fig.~2(c)).  We identify the field-magnitude-independent part as due to the spin Nernst effect. 
For the $l = 1.8$ mm device with a  heater power of 442 mW ($l\nabla$T = 6.8 $\pm$ 0.1 K) we find $S_{SNT}\times l\nabla T = 610~\pm~70$ nV. 
Figure 3(c) shows (as expected) that the cos$^2\theta$ term of the YZ magneto-thermopower varies linearly with temperature difference across the device (i.e., for different heater powers).

We have checked that the signals corresponding to the spin Nernst thermopower are absent for single layers of either Bi$_2$Se$_3$ or CoFeB.  The magneto-thermopower of Bi$_2$Se$_3$ by itself (Fig.~4(a)) has a $\cos^2\theta$ contribution, but 
with a field-magnitude-independent part that would be negligible for the bilayer after accounting for shunting.  
The magneto-thermopower of the CoFeB layer by itself (Fig.~4(b)) contains only a $\cos\theta$ dependence to measurement accuracy, corresponding to the anomalous Nernst voltage with no spin Nernst signal.

To convert our measurement of $S_{SNT}$ to the spin Nernst ratio $\theta_{SN}$  using Eq.~\ref{eq_S_SNT} we must also determine the absolute Seebeck coefficient $S_{SS}$ and the spin Hall ratio $\theta_{SH}$ of the Bi$_2$Se$_3$ spin source. To measure the Seebeck coefficient of Bi$_2$Se$_3$, we grew a layer of Bi$_2$Se$_3$ (8 nm) capped with 20 nm of Se, removed the Se with the same process used for the Bi$_2$Se$_3$ samples, and then grew 2 nm of Ta which oxidizes upon air exposure. The Seebeck coefficient was measured using a similar sample geometry as the spin Nernst magneto-thermopower (Fig.~\ref{Fig1}(d)), but in the absence of a magnetic field.  We obtain a Seebeck coefficent for 8 nm Bi$_2$Se$_3$ relative to the Ti(5 nm)/Pt(70 nm) electrodes to be -87 $\frac{\mu V}{K}$ (Fig.~3(d)). The absolute Seebeck coefficient of the electrodes should be small ($\approx$ -1 $\frac{\mu V}{K}$~\cite{kockert2019absolute}), so we estimate that the absolute Seebeck coefficient of the 8 nm Bi$_2$Se$_3$ is $S_{SS} = - 86$ $\frac{\mu V}{K}$ with a few percent uncertainty.
The measured Seebeck coefficient $S$ of the full bilayer is quantitatively consistent using the measured value $S_{SS}$ for the isolated Bi$_2$Se$_3$ layer and the estimated shunting parameter $\chi$ of the bilayer (see supplemental material).

Given our measurements that $S_{SNT} = 90 \pm 10$ $\frac{nV}{K}$, $S_{SS} = - 86$ $\frac{\mu V}{K}$, and $\Delta R_{SMR}/R = a/100 = (1.26 \pm 0.08) \times 10^{-3}$, Eq.~\ref{eq_S_SNT} yields the ratio of the spin Nernst ratio to the spin Hall ratio to be $\theta_{SN}/\theta_{SH} = -0.83 \pm 0.15$.  By Eq.~\ref{eq_bareSN2} the corresponding ratio for the bare spin Nernst ratio is $\theta^0_{SN}/\theta_{SH} = -0.67 \pm 0.15$. 
Values of order 1 for the ratios $\theta_{SN}/\theta_{SH}$ and $\theta^0_{SN}/\theta_{SH}$ have also been measured for heavy metals~\cite{sheng2017spin, meyer2017observation, kim2017observation, bose2022origin}.

    To determine the values of $J_s/\nabla_x T$, $\theta_{SN}$, and $\theta^0_{SN}$ by themselves, we must also know the spin Hall ratio $\theta_{SH}$ for the Bi$_2$Se$_3$ layer.  To estimate this quantity,  we performed a second harmonic Hall analysis~\cite{hayashi2014quantitative} for the Bi$_2$Se$_3$/CoFeB bilayer 
    (see the supplementary information).   This analysis yields a spin torque efficiency $\xi_{DL}$ that is related to $\theta_{SH}$ according to $\xi_{DL} = T_\text{int} \theta_{SH}$, where $T_\text{int} \le 1$ is an interfacial spin transparency factor.  We find $\xi_{DL} = 0.73$ $\pm$ $0.02$, which therefore provides a lower bound for $\theta_{SH}$.  
    We conclude that a lower bound for $J_s/\nabla_x T$ is (4.9 $\pm$ 0.9)  $\times$ 10$^{6}$ ($\hbar/2e$)  A~m$^{-2}$ / K $\mu$m$^{-1}$, that a lower bound for the magnitude of  $\theta_{SN}$ 
    is approximately  $- 0.61 \pm 0.11$, and that a lower bound for the magnitude of  $\theta^0_{SN}$
    is  -0.49 $\pm$ 0.09.
   Comparisons to previous measurements of heavy metals are provided in Table \ref{table1_comp}. 

\section{Conclusions}

In summary, by comparing measurements of the spin Nernst magneto-thermopower in Bi$_2$Se$_3$/CoFeB bilayers to the spin Hall magnetoresistance, we find a lower bound for the magnitude of the spin Nernst ratio for Bi$_2$Se$_3$ of $-0.61~\pm~0.11$, roughly 3 times greater than that of previously measured values for Pt and 2-3 times greater than that of W. Moreover, the net spin current generated per unit thermal gradient $J_s/\Delta T_x$ is higher in Bi$_2$Se$_3$ than for W and of a similar magnitude as Pt despite the higher resistivity of Bi$_2$Se$_3$.
Interestingly, although the spin Nernst ratio in Bi$_2$Se$_3$ is enhanced relative to the heavy metals, the degree of enhancement appears to be less than for the corresponding spin Hall ratios.

Finally, we address an important question regarding these enhancements in charge/spin and thermal-gradient/spin conversion efficiency: do the topological surface states play a role? The short spin diffusion length in Bi$_2$Se$_3$~\cite{wang2016surface} suggests that states at the surface must be involved. However, we note that there is currently no known experimental technique that can ascertain whether the topological nature of the surface states shown in Fig.~1(c) survives after the deposition of a metallic ferromagnetic overlayer. Indeed, first principles calculations indicate that the surface states at such interfaces are likely to be complicated~\cite{zhang2016band, hsu2017hybridization}. Nonetheless, recent studies have shown that the the density of Berry curvature in the bulk band structure of topological insulators such as Bi$_2$Se$_3$ leads to a large spin Hall conductivity. Bulk-surface correspondence then implies that efficient spin-charge interconversion will also occur via surface states~\cite{wang2019fermi}.

\section{Acknowledgements}
We thank Steve Kriske for useful suggestions on the experiment and Matthias Althammer and Gaël Grissonnanche
for helpful discussions. R.J. grew, fabricated and measured thin films under the supervision of D.C.R with assistance from A.B.; M.S, A.R.R. and T.P. grew the Bi$_2$Se$_3$ films under the guidance of N.S.; X.S.Z. performed Cross-sectional scanning transmission microscopy under the supervision of D.A.M.; R.J and D.C.R. wrote the manuscript with feedback from all authors.

This work made use of the Cornell Center for Materials Research Shared Facilities, which are supported through the NSF MRSEC program (DMR-1719875), and was
 performed in part at the Cornell NanoScale Facility, a member of the National Nanotechnology Coordinated Infrastructure (NNCI), which is supported by the National Science Foundation (Grant NNCI-2025233). R.J.\ was supported by US Department of Energy (DE-SC0017671) and A.B.\ was supported in part by the DOE and in part by the NSF through the Cornell Center for Materials Research (DMR-1719875).  We also acknowledge the financial support of the National Science Foundation through the Penn State 2D Crystal Consortium—Materials Innovation Platform (2DCC-MIP) under NSF cooperative agreement DMR-1539916 and DMR-2039351.  D.A.M and X.S.Z acknowledge  NSF through the Cornell Center for Materials Research (DMR-1719875).

\bibliography{main}
\newpage

\begin{table*}[!]
    \centering
    \begin{tabular}{| *{7}{c|}}
    \hline
 & Bi$_2$Se$_3$& \multicolumn{3}{c|}{W} & \multicolumn{2}{c|}{Pt} \\
\hline
& This work & ~\cite{bose2022origin} & ~\cite{sheng2017spin} & ~\cite{kim2017observation} & ~\cite{meyer2017observation} & ~\cite{kim2017observation}\\
\hline
$\theta_{SN}$ & $-0.61$ $\pm$ 0.11 & - & 0.2 & 0.32 $\pm$ 0.1 &  $-$0.2 & $-$0.18 $\pm$ 0.06\\
\hline
$ \frac{J_{s}}{\nabla T}$ (A~m$^{-2}$/K~$\mu$m$^{-1}$)& (4.9 $\pm$ 0.9)  $\times$ 10$^{6}$ & - & 1.84 $\times$ 10$^{6}$  & (2.6 $\pm$ 0.8) $\times$ 10$^{6}$ & 1.21 $\times$ 10$^{6}$ & (6 $\pm$ 2) $\times$ 10$^{6}$ \\
 \hline
 $ \frac{\theta_{SN}}{\theta_{SH}}$ & $- 0.83$ $\pm$ 0.15   & $-$2.4 $\pm$ 0.6& $-$0.7 & $-$1.53 $\pm$ 0.47& $-$0.6 &$-$1.8 $\pm$ 0.6\\
 \hline
 $\xi_{DL}$ &0.73 & - & $-0.28$ & $-$0.21 & 0.11 & 0.1\\
\hline
S$_{SS}$ ($\frac{\mu V}{K}$) &$-86.6$ & $-$4.5 & $-$12& $10$ & $-$2.6 & $-$10\\
\hline 
$\rho_{SS}$ ($\mu \Omega cm$) & 1064 & 204 & 130 & 125 & 43 & 30\\
 \hline
    \end{tabular}   
    \caption{Comparison of the different figures of merit for the bilayers studied in this work with the existing literature. S$_{SS}$ was measured in ~\cite{bose2022origin, sheng2017spin} and calculated in~\cite{meyer2017observation}, while ~\cite{kim2017observation} referenced existing literature for W and Pt samples. Values for $\frac{J_{s}}{\nabla T}$ for previous studies are calculated using parameters in the table.}
    \label{table1_comp}
\end{table*}

\newpage

\newpage
\newpage
\begin{figure}[h]
    \centering
    \includegraphics[width = \linewidth]{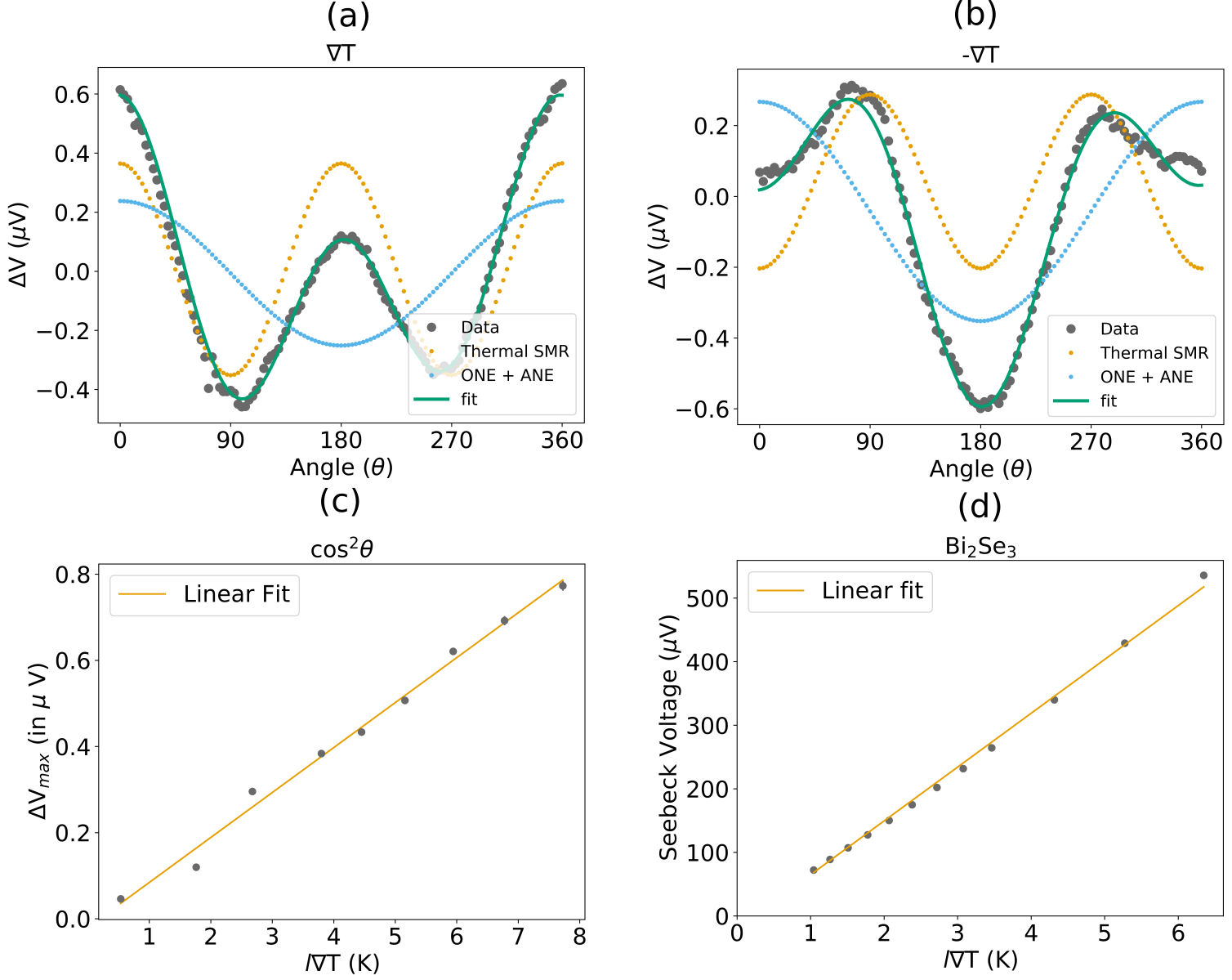}
    \caption{(a,b) YZ magneto-thermopower for a field magnitude of 4 Tesla and a heater power of 335 mW (corresponding to a temperature drop of 5.16 $\pm$ 0.05 K across the device) for (a) one sign of in-plane temperature gradient and (b) the opposite in-plane thermal gradient.  
    (c) Dependence of the cos$^2\theta$ signal on the temperature difference across the device.
    (d) Seebeck voltage for an 8 nm Bi$_2$Se$_3$ layer as a function of the temperature difference across the device. 
    The Seebeck coefficient relative to the Ti/Pt electrodes was determined to be -87 $ \pm$ 1 $\mu$V/K consistent with previous studies~\cite{guo2016tuning}. 
    }
\end{figure}

\newpage
\begin{figure}[h]
    \centering
    \includegraphics[width = \linewidth]{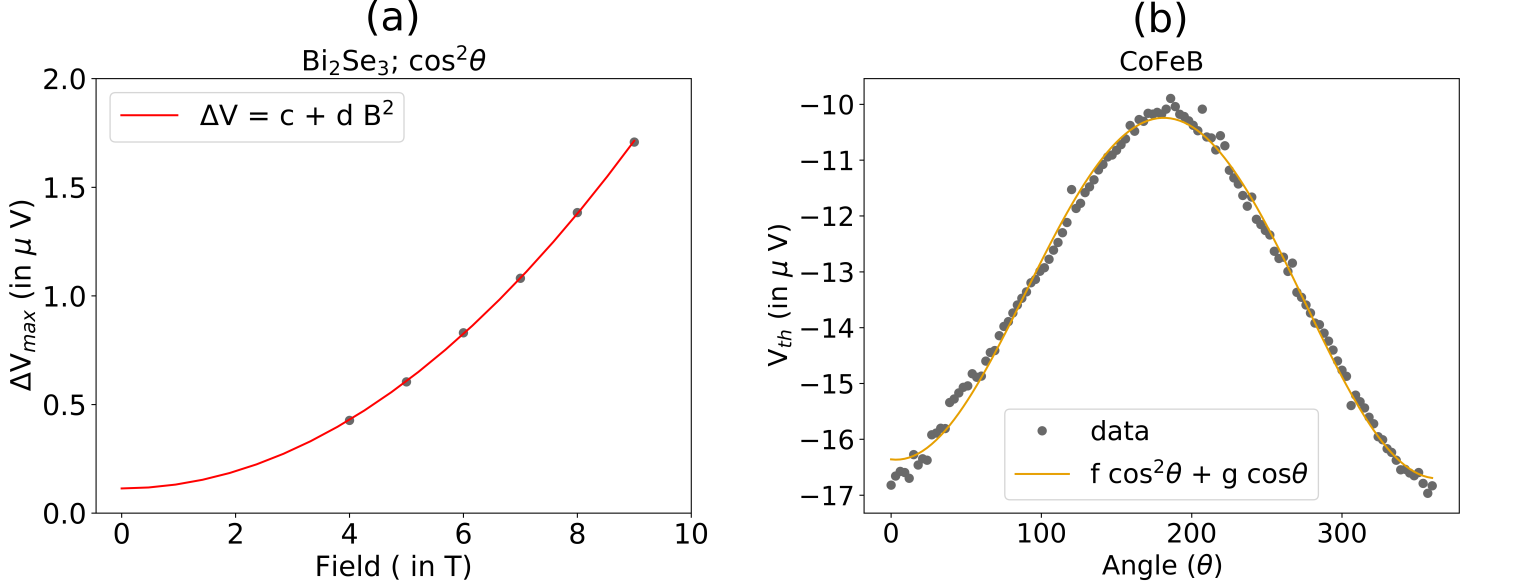}
    \caption{Comparisons which show negligible contributions to the spin Nernst magneto-thermopower in single layers of Bi$_2$Se$_3$ and CoFeB. (a) Field dependence of the $\cos^2\theta$ component of the YZ magneto-thermopower from a 8 nm Bi$_2$Se$_3$ film.  The thermal difference was fixed at 2.17 K $\pm$ 0.02 K.  The is fit to the form $c + d B^2$  with $c$ = 113 nV $\pm$ 25 nV and $d$ = 0.020 $\mu$V/T$^2$. (The contribution in a bilayer after accounting for shunting would therefore be just $\frac{c \chi}{\Delta T}$ = 8.3 $\pm$ 1.8 nV/K.) (b) Angle dependence of the YZ magneto-thermopower signal for a 5 nm CoFeB film at 5 Tesla for the heater power of 318 mW. The fit is to the form $ f \cos^2\theta + g\cos\theta$ with $f$ = 0.04  $\pm$ 0.02 $\mu$V (indicating a negligible spin Nernst effect) and $g$ = $-$3.14 $\pm$ 0.02 $\mu$V.
    }
\end{figure}

\end{document}


\preprint{APS/123-QED}

\title{SI: Thermally-generated spin current in the topological insulator Bi$_2$Se$_3$}
\author{Rakshit Jain}
\affiliation{Department of Physics, Cornell University. Ithaca, NY 14853, USA}
\affiliation{ School of Applied and Engineering Physics, Cornell University. Ithaca, NY 14853, USA}
\author{Max Stanley}
\affiliation{Department of Physics and Materials Research Institute, The Pennsylvania State University, University Park, 16802, Pennsylvania, USA}
\author{Arnab Bose}
\affiliation{Department of Physics, Cornell University. Ithaca, NY 14853, USA}
\affiliation{ School of Applied and Engineering Physics, Cornell University. Ithaca, NY 14853, USA}
\author{Anthony R. Richardella}
\affiliation{Department of Physics and Materials Research Institute, The Pennsylvania State University, University Park, 16802, Pennsylvania, USA}
\author{Xiyue S. Zhang}
\affiliation{ School of Applied and Engineering Physics, Cornell University. Ithaca, NY 14853, USA}
\author{Timothy Pillsbury}
\affiliation{Department of Physics and Materials Research Institute, The Pennsylvania State University, University Park, 16802, Pennsylvania, USA}
\author{David A. Muller}
\affiliation{ School of Applied and Engineering Physics, Cornell University. Ithaca, NY 14853, USA}
\affiliation{Kavli Institute at Cornell for NanoScale Science, Ithaca, NY 14853, USA}
\author{Nitin Samarth}
\affiliation{Department of Physics and Materials Research Institute, The Pennsylvania State University, University Park, 16802, Pennsylvania, USA}
\author{Daniel C. Ralph}
\affiliation{Department of Physics, Cornell University. Ithaca, NY 14853, USA}
\affiliation{Kavli Institute at Cornell for NanoScale Science, Ithaca, NY 14853, USA}

\date{\today}
\maketitle
\section{Thin-film Growth}
The Bi$_2$Se$_3$ films were grown by molecular beam epitaxy (MBE) in a ScientaOmicron EVO50 system. Sapphire (0001) substrates were initially outgassed in vacuum at roughly 600 °C and were then cooled to a growth temperature of 220 °C, as measured by a pyrometer with an emissivity of 0.7. Bismuth (99.999\%) and selenium (99.999\%) from Alfa Aesar were co-evaporated with a flux ratio of 1:10 until a thickness of 8 nm was achieved. The Bi$_2$Se$_3$ films were then capped with 20 nm of Se. The Se capping layers are amorphous and are deposited with a manipulator (substrate) temperature of 10 °C, with the Se crucible at 210 °C and the Se cracker at 350 °C. The film thicknesses were calibrated using both atomic force microscopy and x-ray reflectivity. 

After the samples were transferred through air to a separate vacuum system for deposition of the magnetic layer, the films are heated to 285 $^{\circ}$C for 3.5 hours to remove the selenium cap. Sputter deposition (30 W in an argon pressure of 2 mTorr) was then used to grow layers of Co$_{20}$Fe$_{60}$B$_{20}$ (5 nm) and Ta (1.2 nm).  The purpose of the Ta cap is to protect the CoFeB layer from oxidizing.

\section{Materials characterization of the B\MakeLowercase{i}$_2$S\MakeLowercase{e}$_3$ layers}
See Fig.~\ref{fig_S1}.

\section{Characterization of the B\MakeLowercase{i}$_2$S\MakeLowercase{e}$_3$ Layers by ARPES}
After growth, the band structure of the Bi$_2$Se$_3$ films (with no capping layer) was measured using in-situ ARPES, performed with the 21.2 eV helium I $\alpha$ spectral line from a helium plasma lamp and a ScientaOmicron DA 30L analyzer with 6 meV energy resolution.  (See Fig.~1(c) in the main text.) 

\section{Device Fabrication and Layer Resistivities}
Hall-bar devices were used for measuring both the magnetoresistance and magneto-thermopower of the Bi$_2$Se$_3$/CoFeB devices.  These were fabricated using three-step photolithography. First, the Hall bars themselves (2500 $\mu$m long by 200 $\mu$m wide with 9 pairs of Hall contacts (only 5 are depicted in the Fig.~1(d) in the main text)) were defined and etched using an Ar-ion plasma. 
Then heaters positioned 15 $\mu$m beyond the each end of the Hall bar were defined using optical lithography followed by sputtering of Ti(5 nm)/Pt($\approx$ 24 nm). In the final step, electrical contacts were defined using lithography followed by sputtering of Ti(5 nm)/Pt($\approx$ 70 nm) electrodes. The masks used for fabrication can be found with the supporting data.  Control samples consisting of single layers of Bi$_2$Se$_3$ and CoFeB were fabricated similarly. 
We measured the magneto-thermopower between the fourth and the ninth Hall contacts away from a heater in order to reduce the unintentional out of plane thermal gradient present in the devices~\cite{bose2022origin}.

The four-point resistivity of an individual 8 nm layer of the Bi$_2$Se$_3$ was measured to be 1064 $\mu \Omega$ cm and the resistivity of an individual 5 nm layer of the CoFeB was measured to be 128 $\mu \Omega$ cm.  The overall resistance of a 2200 $\mu$m long 200 $\mu$m wide channel of the Bi$_2$Se$_3$ bilayer was 2329 $\Omega$.  This compares well to the value expected ($= 2361$ $\Omega$) from adding the conductances of the individual layers.                                                         

\section{Details of the electrical and thermal measurements}
The magnetoresistance and  magneto-thermopower measurements were carried out in Quantum Design Physical Properties measurement System (PPMS) with a horizontal rotator. The PPMS was interfaced with a home-built breakout box and software for external control. The experiments were carried out with the samples at room temperature.

\section{Thermal gradient calibration}
The thermal gradient within the sample Hall bars produced by the on-chip heaters was calibrated using the temperature dependence of the sample resistivity, as shown in Fig.~\ref{fig_S3}. 
The thermal difference was measured between two points A and B on the device (the same points between which the YZ magneto-thermopower is measured). This was done in the following steps: 
\begin{itemize}
    \item First, the changes in resistance for the pairs of side contacts at positions A and B were measured as a function of temperature. (Fig.~\ref{fig_S3}(b)).  These data were used to calibrate the linear temperature coefficient for points A, B; i.e., for each point to determine the linear temperature coefficient $\alpha$ in the expression: $100 \times \frac{\Delta R}{R} = (\alpha (T - T_0$)) where  $T_0$ is the reference temperature (305 K for this calibration). 
    \item Second, the change in resistance as a function of increasing heater power was measured for the pairs of side contacts at points A and B.  Using the values of $\alpha$ noted in the first step, the temperature of points A and B was calculated (Figure.~\ref{fig_S3}(c)), along with the thermal difference between these two points (Figure.~\ref{fig_S3}(d)).  
\end{itemize}

\section{Measurement of spin Hall torque efficiencies}
To determine the spin Hall torque efficiency we performed second harmonic hall (SHH) measurements~\cite{hayashi2014quantitative, macneill2017thickness} on Bi$_2$Se$_3$ (8 nm)/CoFeB (2.6 nm) bilayers.  (We used a thinner CoFeB layer than for the magneto-themopower measurements in the main text in order to enhance the torque signals.)
The bilayers were etched into hall bar shape with width 6 $\mu$m and length 20 $\mu$m, with side contacts 5 $\mu$m  wide.  The device resistance was be 1340 $\Omega$.
We applied a low-frequency alternating current (1317 Hz in this experiment) of I$_{RMS}$ = 3 mA through the device and measured the Hall-voltage response at the 1st and 2nd harmonic frequencies (1317 Hz and 2634 Hz). 
These data are plotted in Fig.~\ref{fig_S4}.

The first and second harmonic response in presence of in plane external field (H) can be fitted to the following expressions (we ignore in-plane anisotropy)~\cite{hayashi2014quantitative} 
\begin{align}
    V_{\omega} &= V_{PHE}~\sin2\phi + V_{AHE}~\cos\theta \\
    V_{2\omega} &=  \frac{{\Delta H_{FL}} V_{PHE}}{H} \cos 2\phi \cos \phi  + \bigg(\frac{{\Delta H_{DL}} V_{AHE}}{2 (H +  H_{\perp})} + V_{ANE} + V_{ONE} H \bigg)\cos \phi,
\end{align}
where $V_{PHE}$ denotes the planar hall voltage, $V_{AHE}$ denotes the anomalous Hall voltage, $V_{ANE}$ denotes the anomalous Nernst voltage, $V_{ONE}$ denotes the ordinary Nernst voltage, and $H_{\perp}$ denotes the field required to saturate CoFeB in the out of plane direction. 
$\Delta H_{FL}$ denotes effective field corresponding to the field-like torque and $\Delta H_{DL}$ denotes the effective field corresponding to the damping-like torque.
From the fits we obtain $\Delta H_{DL}$ to be 16.4 Oe and $\Delta H_{FL}$ to be 1.36 Oe. The contribution from the Oersted field  ($\frac{\mu_0 J_e t_{BS}}{2}$) is estimated to be 0.97 Oe. These data can be converted into damping and field-like torque efficiencies given by 
\begin{align}
    \xi_{DL, FL} = \frac{2 e}{\hbar} \frac{\mu_0 M_s t_{CFB} \Delta H_{DL, FL}}{ J_e}
\end{align}
where $J_e$ denotes the current density through the device, $M_s$ denotes saturation magnetization which is measured to be 1.53/$\mu_0$ T using vibrating sample magnetometry, and $t_{CFB}$ denotes the thickness of the CoFeB magnetic layer.
From this analysis we obtain $\xi_{DL}$ to be 0.73 $\pm$ 0.02 and $\xi_{FL}$ to be 0.02 $\pm$ 0.01.

One potential concern about performing second harmonic Hall measurements of the spin-orbit torque from a topological insulator is that there could be artifacts arising from non-linear Hall contributions unrelated to the spin torque~\cite{yasuda2017current}. However, this effect appears to be important primarily at low temperatures, and has been shown to be have negligible at temperatures above 200 K~\cite{wang2020one}. The measurements we present are limited to room temperature.
\section{Comparison of Seebeck coefficients in  bilayer vs single layer of B\MakeLowercase{i}$_2$S\MakeLowercase{e}$_3$ }
From the background voltage of the measurement in Fig.~2(b), the Seebeck coefficient of the bilayer is determined to be $S= - 13.27 \pm 0.25$ $\frac{\mu V}{K}$.  The resistivity of an individual 8 nm Bi$_2$Se$_3$ layer is 1064 $\mu$Ohm cm and for an individual 5 nm CoFeB  layer is 128 $\mu$Ohm cm. We therefore estimate that the shunting coefficient of the bilayer is $\chi =$ 0.16. After accounting for shunting, we estimate that for the Bi$_3$Se$_3$ layer in the absence of shunting $S_{SS} = S/\chi = - 83 \pm 2$ $\frac{\mu V}{K}$. The close consistency with the value measured directly for the individual Bi$_2$Se$_3$ layer ($-86$ $\frac{\mu V}{K}$, see the main text) suggests that the thermoelectric properties of the Bi$_2$Se$_3$ in the bilayer are not strongly changed by interaction with the CoFeB layer.

\section{YZ Magnetoresistance in single layers of B\MakeLowercase{i}$_2$S\MakeLowercase{e}$_3$ and C\MakeLowercase{o}F\MakeLowercase{e}B}

In order to test for any contribution to the spin Hall magnetoresistance of the bilayer arising from the individual layers of Bi$_2$Se$_3$ and CoFeB, we performed YZ magneto-resistance measurements on single layers of Bi$_2$Se$_3$ (8 nm) and CoFeB (5 nm), both capped with 2 nm of Ta which oxidizes upon exposure to air. The YZ magnetoresistance of Bi$_2$Se$_3$ is plotted in Fig.~\ref{fig_S5}(a) and the field dependence of the cos$^2 \theta$ fits is plotted in Fig.~\ref{fig_S5}(b). A fit of the data in Fig.~\ref{fig_S5}(b) to the form $a + bB^2$ with fitting parameters $a$ and $b$ indicates a significant contribution $\propto B^2$, but the zero-field extrapolation is just $a=0.0020 \pm 0.0006$, which is negligible compared to the SMR signal 0.155 in the main text.  We therefore observe a negligible spin Hall magnetoresistance signal for the single layer of Bi$_2$Se$_3$. 

For CoFeB we carried out similar YZ magnetoresistance measurement, finding the angular dependence  plotted in Fig.~\ref{fig_S5}(c). A fit of the data in Fig.~\ref{fig_S5}(d) to the form $a + bB^2$ with fitting parameters $a$ and $b$  yields $a = 0.034$ $\pm$ 0.003 and $b$ = (4.4 $\pm$ 0.6) $\times$ 10$^{-5}$ T$^{-2}$.  The field-independent part important for this study  ($=a$) is non-zero, but roughly 5 times smaller than the  the SMR signal 0.155 in the main text.
This non-zero value could arise from the asymmetry in our structure (Al$_2$O$_3$/CoFeB/TaO$_{x}$)~\cite{seki2021spin}.  Similar signals have been reported in CoFeB/MgO samples in Fig.~S4(b) of~\cite{sheng2017spin}.  If we assume that the CoFeB signal and the SMR contribute in parallel to the conductance of the bilayer and subtract off the CoFeB signal, our estimate for the SMR signal is $100 \times \Delta R_{SMR}/R =$ 0.126 $\pm$ 0.08.  

We have also compared the magnetic-field dependences of the YZ magneto-thermopower and magnetoresistance between the Bi$_2$Se$_3$/CoFeB bilayer and the individual Bi$_2$Se$_3$ films capped with oxidized Ta.  For the YZ magneto-thermopower, if the field-dependent part of the signal originates entirely within the TI layer we would expect the voltage signal in the Bi$_2$Se$_3$/CoFeB bilayer to be reduced relative to the individual Bi$_2$Se$_3$ layer by shunting, i.e., $\frac{d}{\chi l \nabla T}$ should be similar to the two samples, using for the shunting parameter that $\chi \approx 1$ for the single layer and $\chi = {\rho_{FM}t_{SS}}/({\rho_{FM}t_{SS} + \rho_{SS}t_{FM}}) \approx 0.16$ using symbols as defined in the main text.  We find reasonable agreement with this expectation, with $\frac{d}{\chi l \nabla T}$ equal to 10.7 $\times$ 10$^{-3}$ $\frac{\mu V}{T^2 K}$ for the bilayer and 9.1 $\times$ 10$^{-3}$ $\frac{\mu V}{T^2 K}$ for the individual Bi$_2$Se$_3$.

Figure.~\ref{fig_S6} shows a comparison between the Bi$_2$Se$_3$/CoFeB/TaO$_x$ and Bi$_2$Se$_3$/TaO$_x$ samples for the component of the $\cos^2\theta$ magnetoconductance that depends quadratically on the magnetic field.  We plot the change in conductance, rather than resistance, to allow a direct comparison between the two types of samples without needing to correct for the shunting factor. If the portion of the magnetoresistance corresponding to this signal originiated entirely in the TI layer, we should expect the two curves in Fig.~\ref{fig_S6} to be nearly identical.  Instead, they differ in scale by a factor of about 3.  This is surprising to us given that the corresponding quadratic-in-field component of the magneto-thermopower is more similar between the samples. We speculate that the quadratic-in-field part of the magnetoconductance must be more sensitive to the existence of the TI's interface with the CoFeB layer, possibly due to either charge transfer or spin scattering.

\bibliography{supp}

\newpage

\renewcommand{\thefigure}{S1}
\begin{figure}[h!]
    \centering
    \includegraphics[width=\linewidth]{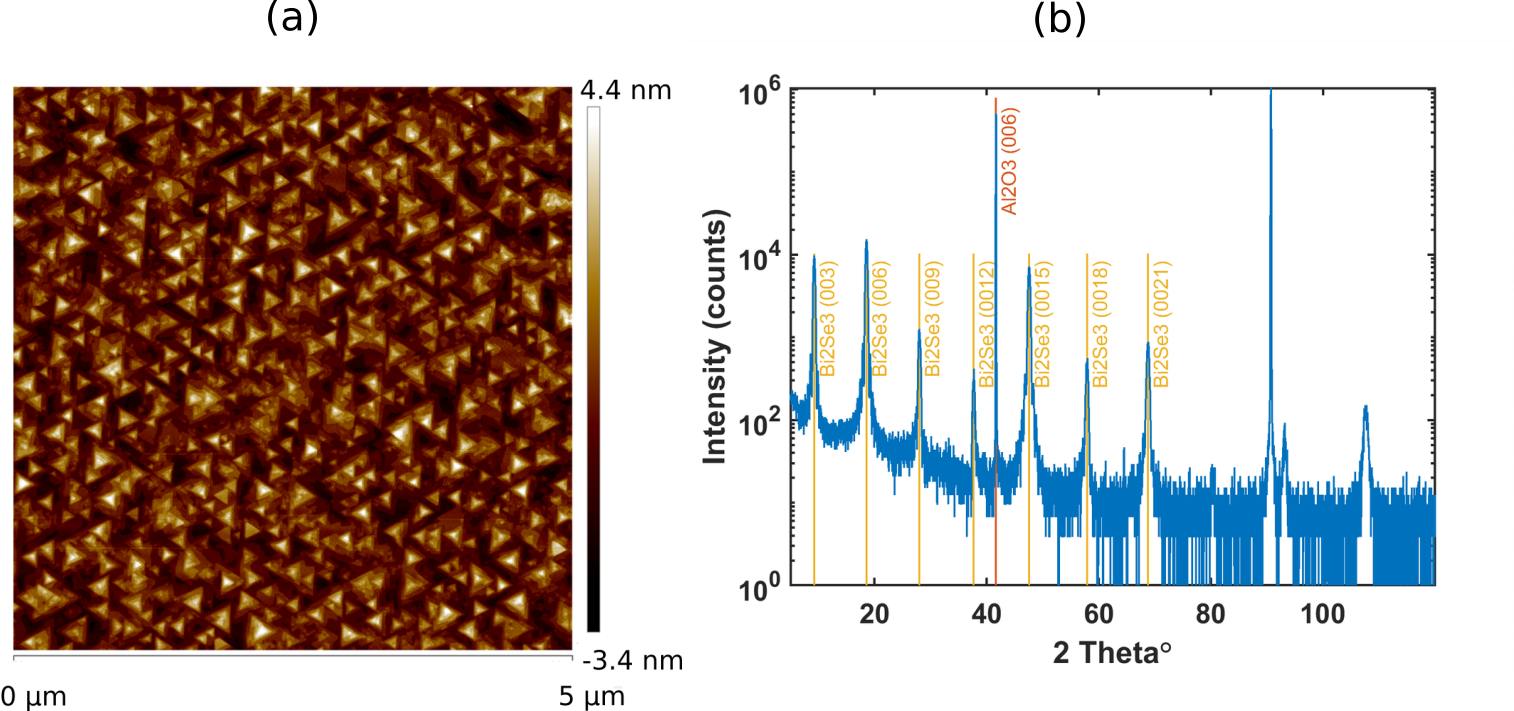}
    \caption{Materials characterization of Bi$_2$Se$_3$ layer (with no Se capping layer). (a) Atomic force microscope image of the Bi$_2$Se$_3$ surface after growth, indicating a root-mean-square roughness of 1.13 nm.  (c) x-ray reflectivity measurement of a Bi$_2$Se$_3$ sample indicating high-quality growth with sharp interfaces.
    }
    \label{fig_S1}
\end{figure}
\renewcommand{\thefigure}{S2}
\begin{figure}[h!]
    \centering
    \includegraphics[width=0.6\linewidth]{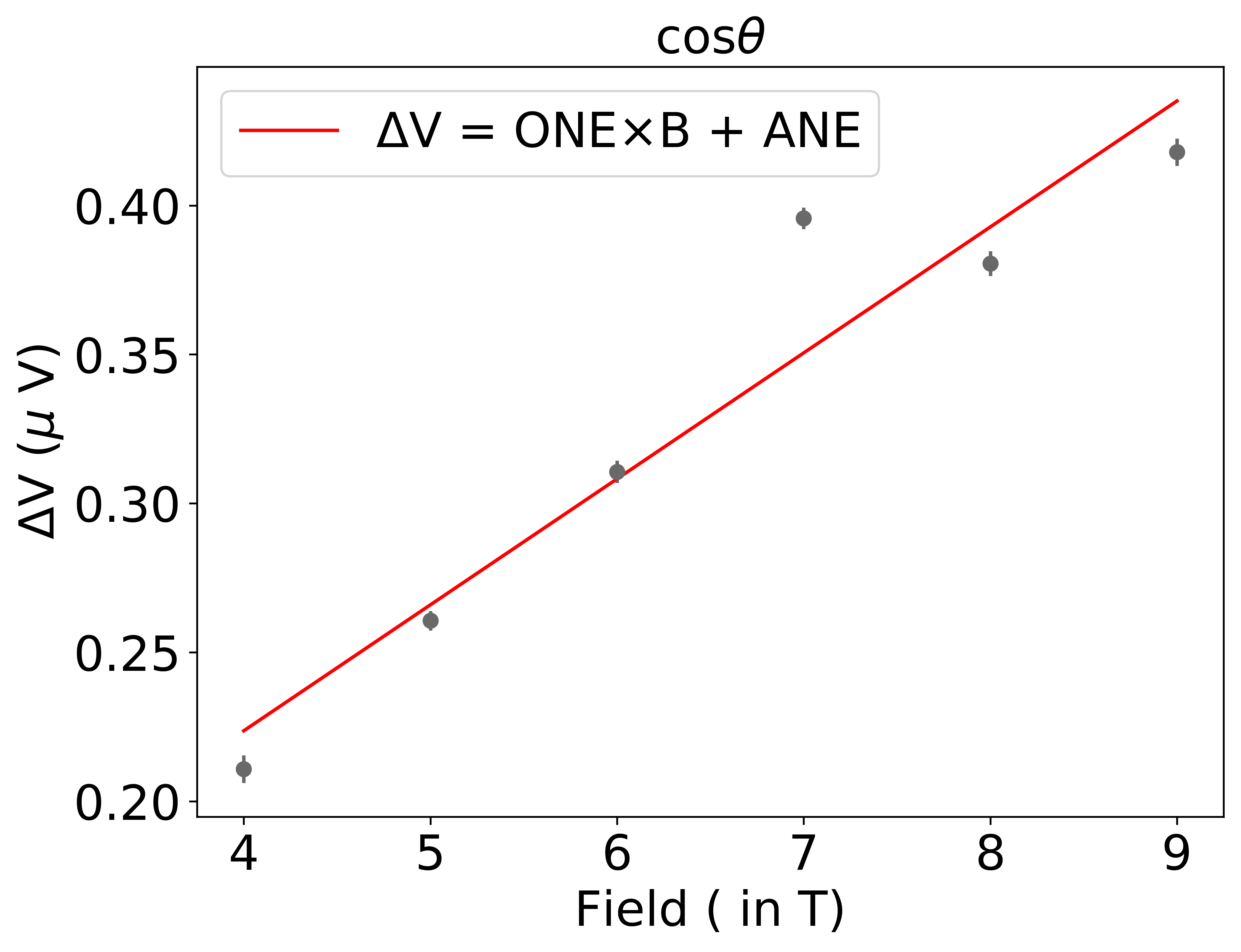}
    \caption{Field dependence of the cos$\theta$ part of YZ magneto-thermopower (Fig.~2(b)) with a fit to the form ONE $\times$ B + ANE with ONE = 0.042 $\pm$ 0.006 $\mu$V/T and ANE = 0.05 $\pm$ 0.04 $\mu$V. ONE and ANE refer to voltages arising from ordinary Nernst effect and anomalous Nerst effect, respectively. }
    \label{fig_S2}
\end{figure}

\renewcommand{\thefigure}{S3}
\begin{figure}[!]
    \centering
    \includegraphics[width = \linewidth]{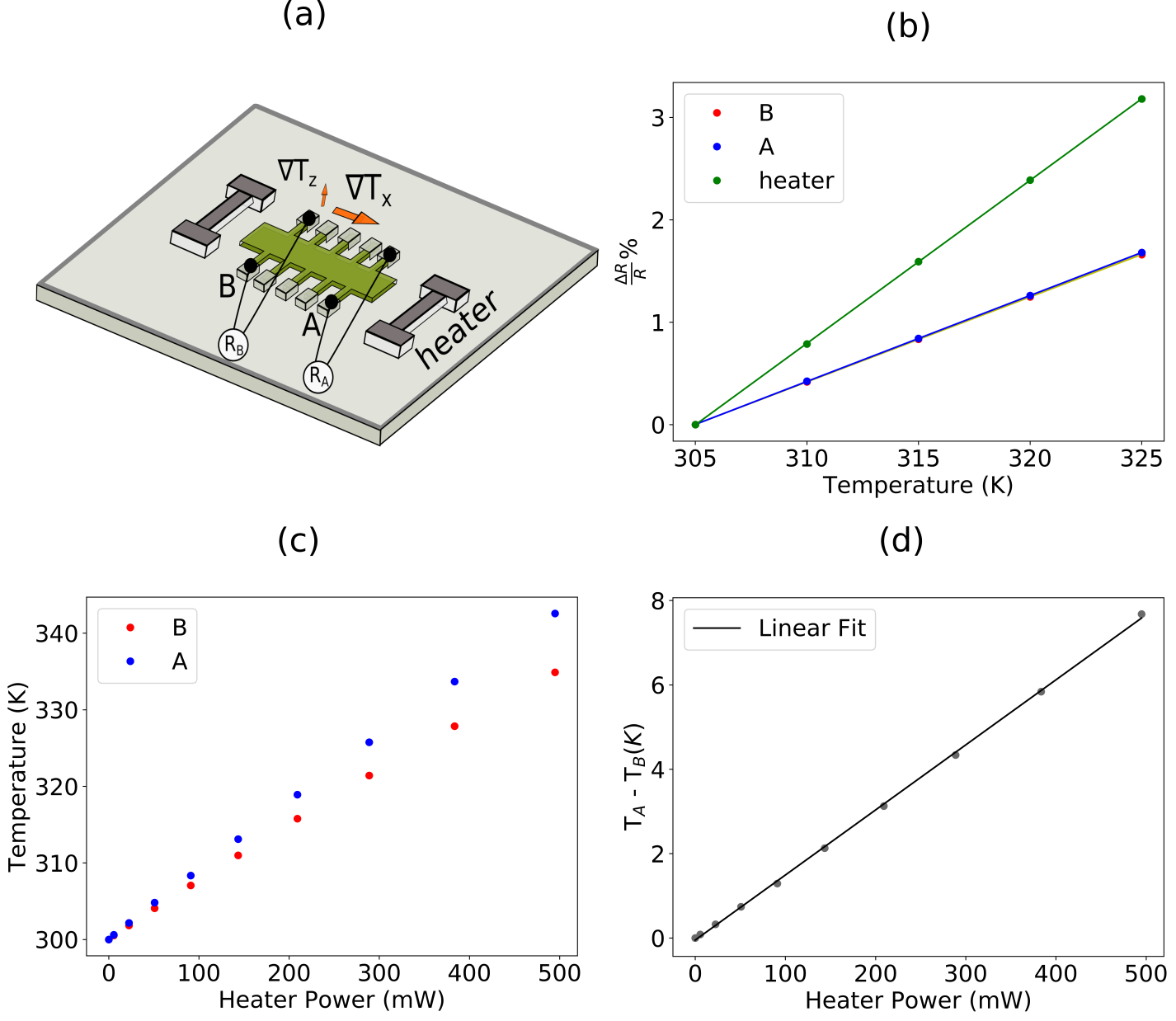}
    \begin{center}
    \end{center}
    \caption{Procedure for calibrating the temperature difference between the points on the sample used for the spin Nernst magneto-thermopower measurement. (a) Sample schematic. (b) Calibration of change in resistance for contacts at points A and B and the heater as a function of sample temperature. (The data for points A and B closely overlap.) We fit the ratios $\frac{\Delta R}{R}$ to a linear function $100 \times \frac{\Delta R}{R} = \alpha (T - T_0$) where $T_0$ is chosen to be 305 K. Linear fits yield $\alpha$ to be 0.083, 0.084, and 0.159 K$^{-1}$ for points A, B, and the heater respectively.
(c) Change in temperature at the points A and B on the sample as the function of heater power.  The temperature changes are determined from the measured changes in resistance, and calibrated using the data in (b). (d) Temperature difference between points A and B as a function of heater power calculated from the difference in (c). }    \label{fig_S3}
\end{figure}

\newpage

\renewcommand{\thefigure}{S4}
\begin{figure}[h!]
    \centering
    \includegraphics[width =\linewidth]{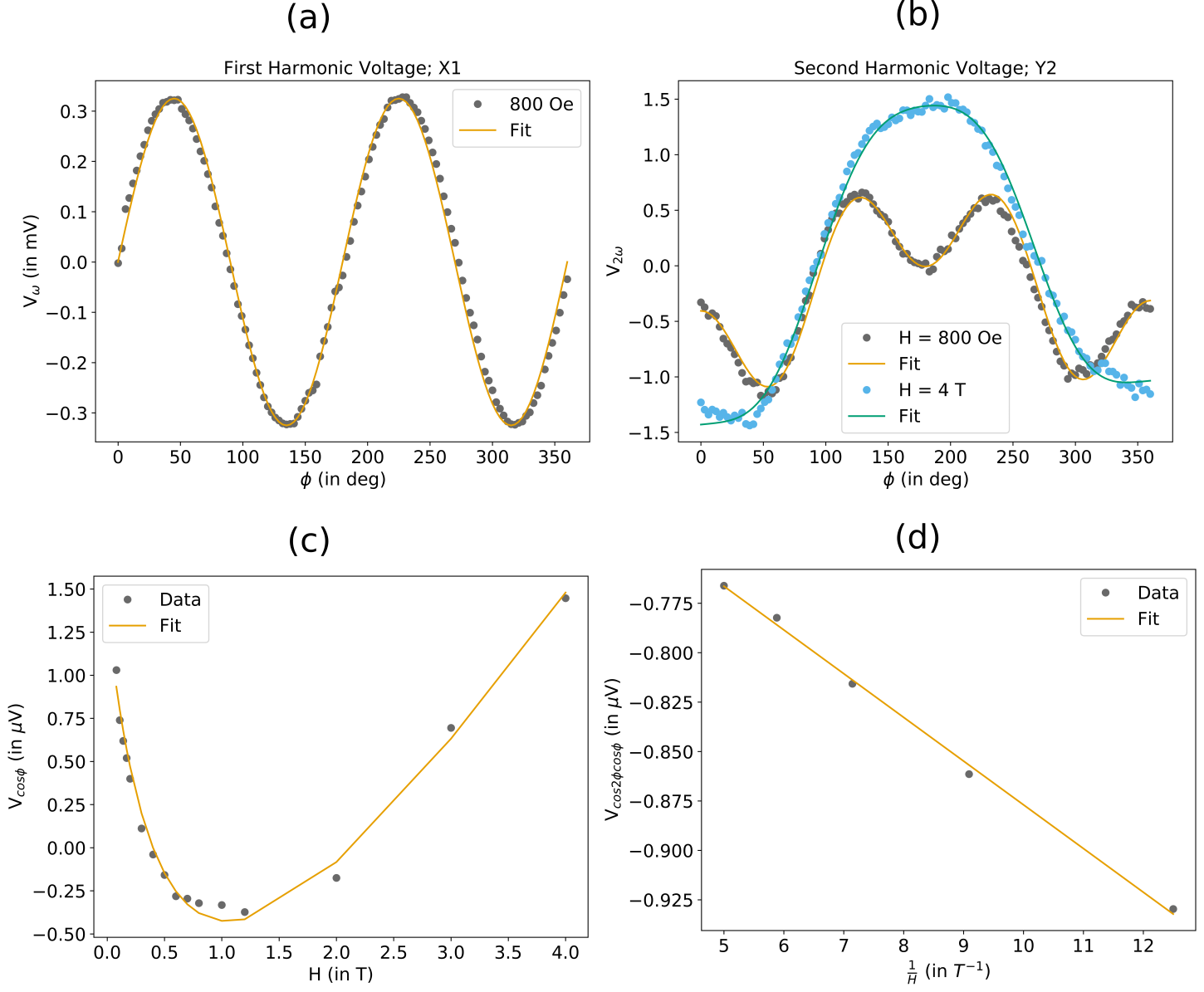}
    \caption{Second harmonic Hall analysis of spin-orbit torques in a bilayer of Bi$_2$Se$_3$(8 nm)/CoFeB(2.6 nm). (a) First harmonic Hall voltage at an external magnetic field of 800 Oe applied in plane as a function of the field angle. (b) Second Harmonic voltages at 800 Oe and 4 T as a function of the angle of the in-plane magentic field. (c) $\cos\phi$ component of the second harmonic Hall voltage versus the magnitude of the applied field. (d) $\cos2\phi \cos\phi$ component of the second harmonic Hall voltage versus the inverse magnitude of the applied field.}
    \label{fig_S4}
\end{figure}

\newpage

\renewcommand{\thefigure}{S5}
\begin{figure}[h!]
    \centering
    \includegraphics[width = \linewidth]{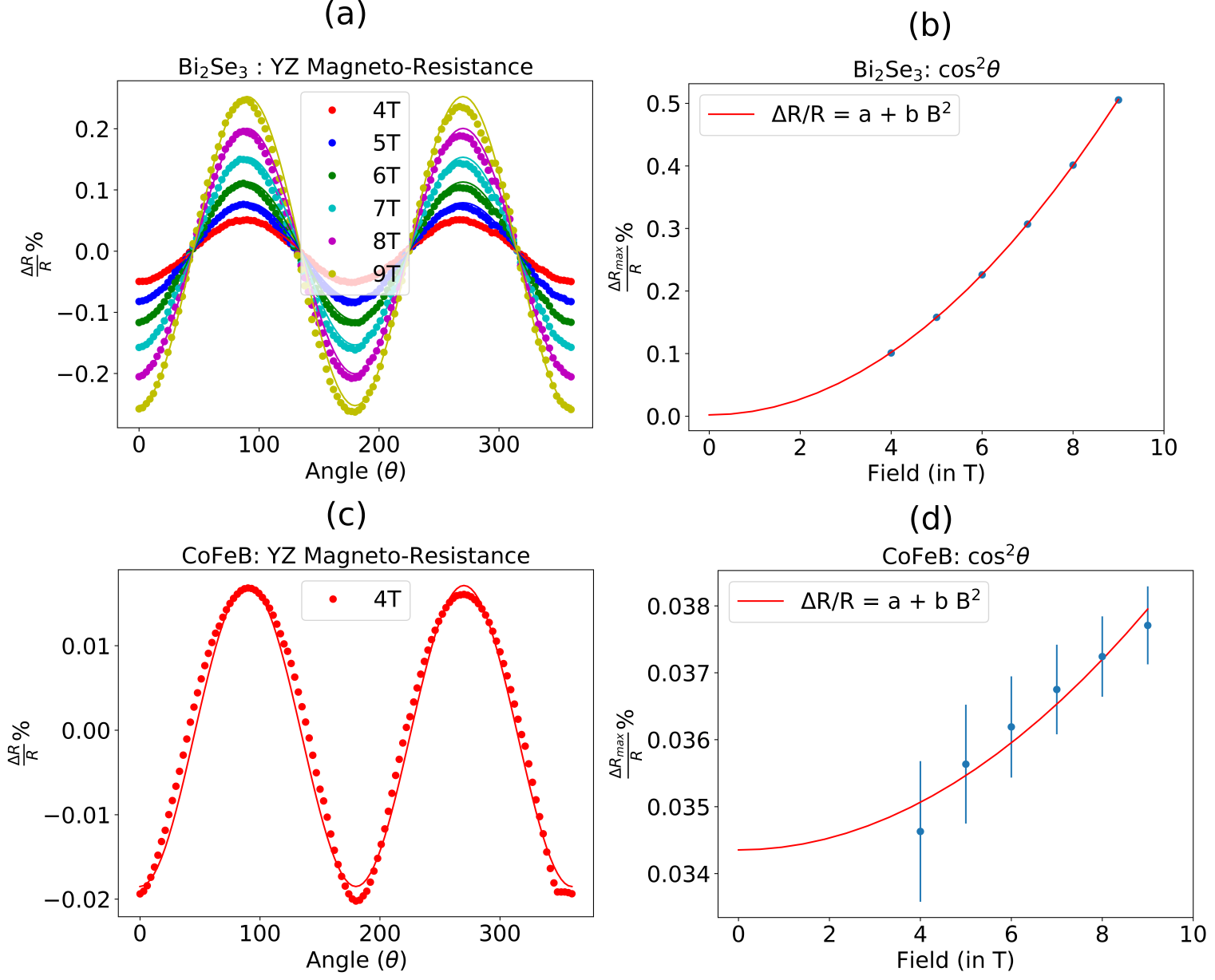}
    \caption{YZ magnetoresistance of single layers of B\MakeLowercase{i}$_2$S\MakeLowercase{e}$_3$ and C\MakeLowercase{o}F\MakeLowercase{e}B. (a) Magnetoresistance percentage ratio ($\frac{\nabla R}{R} \times 100$) as a function of the magnetic field angle and magnitude for Bi$_2$Se$_3$ (8 nm) at room temperature, for magnetic field rotated in the YZ plane.
    (b) Amplitude of the YZ magnetoresistance for Bi$_2$Se$_3$ as a function of magnetic-field magnitude, with a fit to the form $100 \times(a + b B^2)$.   The fit yields $a =0.0020 \pm 0.0006$  and $b = 0.00622\pm 0.00002$ T$^{-2}$. 
   (c) Magnetoresistance percentage ratio ($\frac{\Delta R}{R} \times 100$) as a function of the magnetic field angle and magnitude for CoFeB (5 nm) at room temperature, for magnetic field rotated in the YZ plane.
   (d) Amplitude of the YZ magnetoresistance for CoFeB as a function of magnetic-field magnitude, with a fit to the form $100 \times(a + b B^2)$.   The fit yields $a = 0.034 \pm 0.003$  and $b = (4.4 \pm 0.6) \times 10^{-5}$ T$^{-2}$. 
    }
    \label{fig_S5}
\end{figure}
\newpage
\renewcommand{\thefigure}{S6}
\begin{figure}[h!]
    \centering
    \includegraphics[width = 0.5 \linewidth]{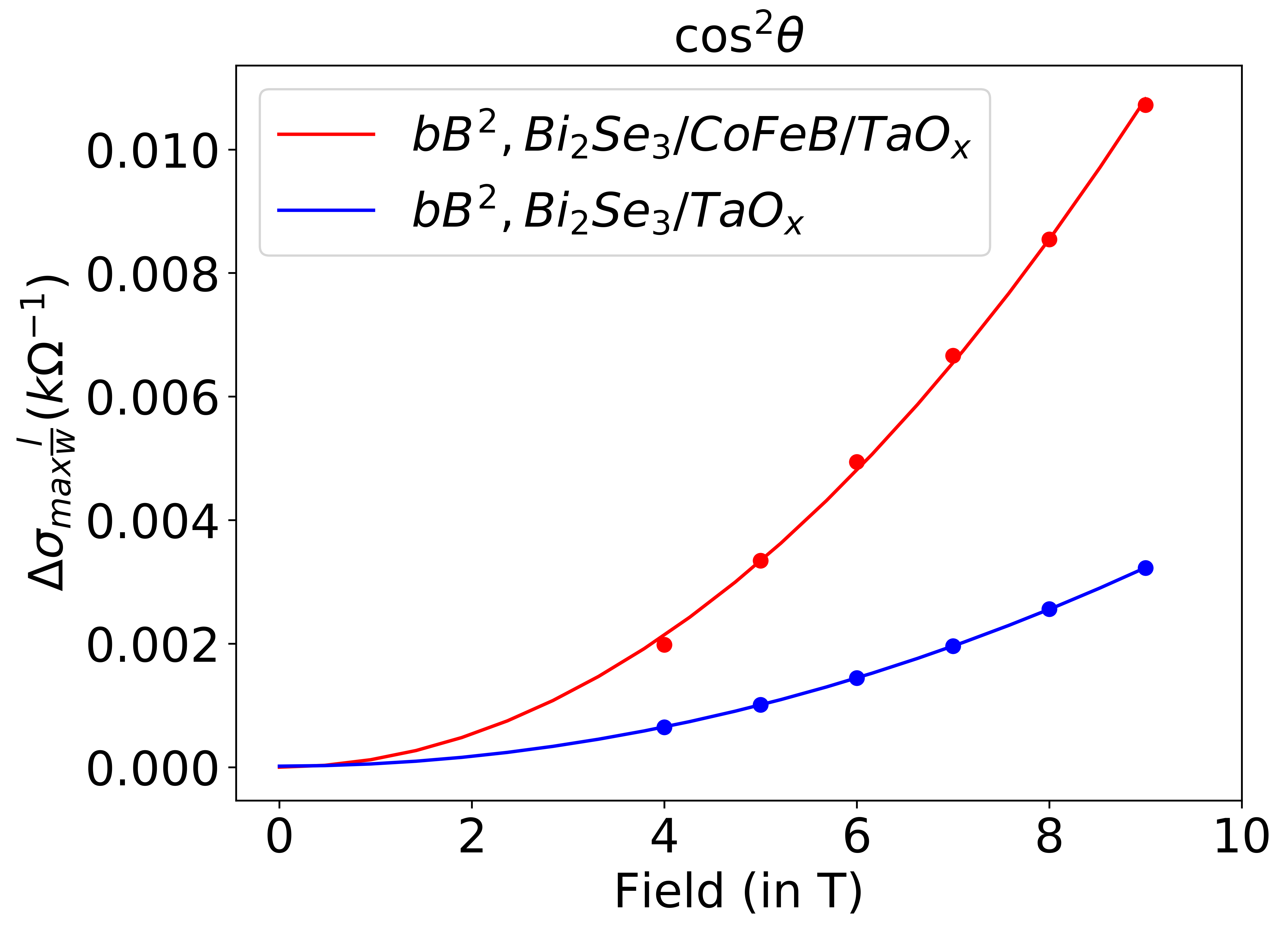}
    \caption{ Comparison of the scaled change in the quadratic-in-magnetic-field portion of the YZ magneto-conductance ($\Delta \sigma_{max} \frac{l}{w}$) for Bi$_2$Se$_3$/TaO$_x$ and Bi$_2$Se$_3$/CoFeB/TaO$_x$, with fits to the form $ b B^2$. (The field-independent part is omitted for clarity.) The fits yields $b = (3.970 \pm 0.008) \times 10^{-5} k \Omega^{-1}$ T$^{-2}$ and $b$ = (13.4 $\pm$ 0.2) $\times$ $10^{-5}$ k$\Omega^{-1}$ T$^{-2}$ respectively for Bi$_2$Se$_3$/TaO$_x$ and Bi$_2$Se$_3$/CoFeB/TaO$_x$.
    }
    \label{fig_S6}
\end{figure}